\documentclass[conference]{IEEEtran}

\IEEEoverridecommandlockouts

\usepackage{cite}
\usepackage{amsmath,amssymb,amsfonts,amsthm}
\usepackage{mathtools}
\usepackage{algorithmic}
\usepackage{graphicx}
\usepackage{textcomp}
\usepackage{xcolor}
\usepackage{algorithm,algorithmic}
\def\BibTeX{{\rm B\kern-.05em{\sc i\kern-.025em b}\kern-.08em
    T\kern-.1667em\lower.7ex\hbox{E}\kern-.125emX}}
\newtheorem{lemma}{Lemma}

\newcommand{\mb}[1]{\mathbf{#1}}
\newcommand{\mcal}[1]{\mathcal{#1}}
\newcommand{\real}[1]{\mathbb{R}^{#1}}
\newcommand{\bt}{\pmb{\theta}}
\newcommand{\ldef}{\stackrel{\Delta}{=}}
\newcommand{\bpsi}{\pmb{\psi}}
\newcommand{\bphi}{\pmb{\phi}}

\newcommand{\norm}[1]{\left\lVert #1 \right\rVert_1}

\DeclareMathOperator{\Ima}{Im}

\begin{document}
\title{Privacy-Preserving Load Forecasting via Personalized Model Obfuscation}

\author{\IEEEauthorblockN{Shourya Bose, Yu Zhang}
\IEEEauthorblockA{\textit{Department of Electrical \& Computer Engineering} \\
\textit{University of California, Santa Cruz, CA}\\
\texttt{\{shbose,zhangy\}@ucsc.edu}}
\and
\IEEEauthorblockN{Kibaek Kim}
\IEEEauthorblockA{\textit{Mathematics and Computer Science Division} \\
\textit{Argonne National Laboratory, Lemont, IL}}
\texttt{kimk@anl.gov}
\thanks{This material is based upon work supported by the U.S. Department of Energy, Office of Science, under contract number DE-AC02-06CH11357. This work was partially supported by the 2023 CITRIS Interdisciplinary Innovation Program (I2P). We gratefully acknowledge the computing resources provided on Swing, a high-performance computing cluster operated by the Laboratory Computing Resource Center at Argonne National Laboratory, and the Hummingbird cluster at UC Santa Cruz.}}

\maketitle

\begin{abstract}
The widespread adoption of smart meters provides access to detailed and localized load consumption data, suitable for training building-level load forecasting models. To mitigate privacy concerns stemming from model-induced data leakage, federated learning (FL) has been proposed. This paper addresses the performance challenges of short-term load forecasting models trained with FL on heterogeneous data, emphasizing privacy preservation through model obfuscation. Our proposed algorithm, Privacy Preserving Federated Learning (PPFL), incorporates personalization layers for localized training at each smart meter. Additionally, we employ a differentially private mechanism to safeguard against data leakage from shared layers. Simulations on the NREL ComStock dataset corroborate the effectiveness of our approach.
\end{abstract}


\section{Introduction}
Efficient and reliable operations of smart power grids rely on accurately forecasting load demand by different consumers~\cite{ME-etal:2007}. 
Yet, the acquisition of detailed load data from smart meters raises privacy concerns for occupants of serviced buildings or residences~\cite{MRA-etal:2017}. To address this, recent research has focused on finding a balance between privacy and accuracy in load forecasting models, particularly those utilizing deep learning architectures.

A promising approach is federated learning (FL)~\cite{AT-SC:2020,MNF-KG-SM:2022}, where data stays local to edge devices (referred to as "clients"), such as smart meters. Meanwhile, model parameters are distributed between the clients and a central server during the training process. For clients with highly diverse data, the performance of federated learning (FL) algorithms tends to decline~\cite{ZC-etal:2019}. Additionally, FL alone may not offer robust privacy guarantees, particularly when models trained through FL exhibit bias toward a single or a group of clients~\cite{AA-etal:2020}. As shown in our prior work~\cite{SB-KK:2023}, the inclusion of personalization layers (PLs) effectively mitigates performance degradation with heterogeneous clients. PLs denote model layers specific to each client, offering an additional layer of privacy protection. This paper delves into the application of PLs in conjunction with a privacy-preserving technique.

In FL training, client model parameters are aggregated at the server, distributed back to clients for local updates on their data. Personalization layers (PL) mark specific model layers for individual clients, not participating in communication. This allows personalized client models, leveraging a shared representation from non-personalized layers~\cite{FEDREP}. PL parameter localization facilitates client model obfuscation during FL training for privacy, with inherent obfuscation for PL parameters and differential privacy (DP) applied to shared layers to prevent data leakages~\cite{CD:2008,MA-etal:2016}.

\subsubsection*{Contribution}
\label{sec:arch}

We introduce a privacy-preserving approach for training short-term load forecasting (STLF) models using heterogeneous clients' load data.The proposed method PPFL (Privacy Preserving Federated Learning) allows for the implementation of PLs while protecting shared layer parameters through DP. It is applicable for training any STLF model. We employ a dual-stage attention-based recurrent neural network (DARNN)~\cite{DARNN} as our chosen learning model in this paper. 
DARNN belongs to attention-based encoder-decoder RNNs, a category that has demonstrated state-of-the-art performance on various load forecasting datasets~\cite{ATTN1,ATTN2,ATTN3}. Simulations reveal that without DP, PPFL-trained models surpass classical FL and local training but exhibit a gradual performance decline with increasing DP levels.

\subsubsection*{Notation}
The set of real numbers is denoted by $\mathbb{R}$.
Vectors are represented in boldface and scalars in standard font.
For vectors $\mb{a}$ and $\mb{b}$, $\mb{a}/\mb{b}$ and $\mb{a}\odot\mb{b}$ denote elementwise division and Hadamard product, respectively. $\mb{a}^{(2)}$ and $\sqrt{\mb{a}}$ denote elementwise square and square root of $\mb{a}$ while $\norm{\mb{a}}$ is the $\ell_1$-norm. $|\mcal{S}|$ represents the cardinality of a finite set $\mcal{S}$. For a positive integer $K$, $[K]$ denotes the set $\{1,\cdots,K\}$.
$\text{Pr}\left[ X\in\mcal{A} \right]$ is the probability that random variable $X$ belongs to set $\mcal{A}$. A Laplace random variable $X\sim\text{Laplace}\left(\mu,b\right)$ has the density function $f(x) = \frac{1}{2b}\exp \left( -\frac{|x-\mu|}{b} \right)$.

\section{Load Forecasting Model Architecture}

\begin{figure}[t]
	\centering
	\includegraphics[width=\linewidth]{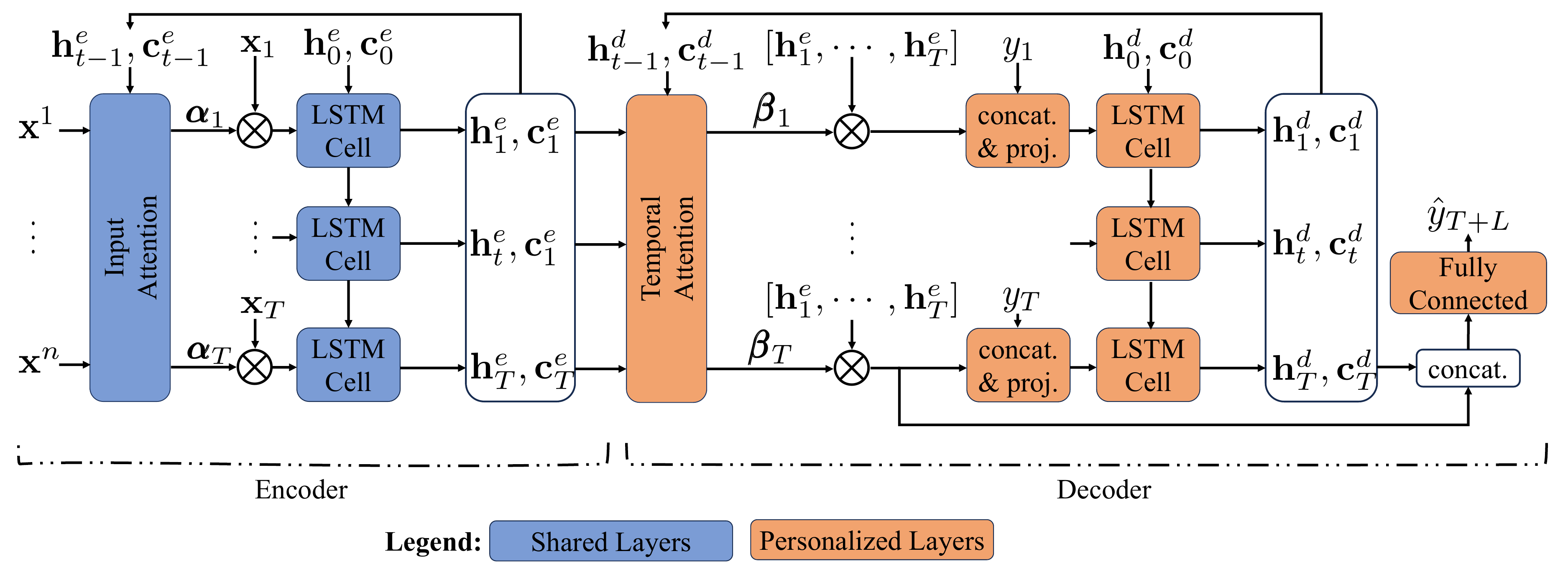}
	\caption{DARNN model architecture for STLF. The shared and personalized layers for algorithms using PLs are marked herein.}
	\label{fig:arch}
\end{figure}

We consider the STLF setup as follows. Given $T$ previous exogenous inputs $\{\mb{x}_t\}_{t=1}^T$ with $\mb{x}_t=[x_{t,1},\dots,x_{t,n}]^\top\in\real{n}$ and loads $[y_1,\dots,y_T] \in\real{T}$, we aim to generate a forecast $\hat{y}_{T+L}$ for some horizon length $L$. The exogenous inputs may contain quantities such as date/time indices, building characteristics, and weather data. In this setting, the LSTM architecture employs a state-based strategy to produce upcoming predictions progressively. The LSTM cell functions as a model for state transitions unfolding over a specified time span. It operates by updating its hidden states $\mb{h}_t$ and cell states $\mb{c}_t$ using input $\mb{x}_t$ and previous states, given as
$\{\mb{h}_t,\mb{c}_t\} = \texttt{LSTMCell}(\mb{x}_t,\mb{h}_{t-1},\mb{c}_{t-1})$.
The internal operations of each LSTM cell are given as follows:
\begin{align*}
	\mb{f}_t &= \sigma(\mb{W}_{fx}\mb{x}_{t}+\mb{W}_{fh}\mb{h}_{t-1}+\mb{b}_f)\\
	\mb{i}_i &=  \sigma(\mb{W}_{ix}\mb{x}_{t}+\mb{W}_{ih}\mb{h}_{t-1}+\mb{b}_i)\\
	\mb{g}_t &=  tanh(\mb{W}_{gx}\mb{x}_{t}+\mb{W}_{gh}\mb{h}_{t-1}+\mb{b}_g)\\	
	\mb{o}_t &=  \sigma(\mb{W}_{ox}\mb{x}_{t}+\mb{W}_{oh}\mb{h}_{t-1}+\mb{b}_o)\\	
	\mb{c}_t &= \mb{g}_t \odot \mb{i}_t + \mb{c}_{t-1} \odot \mb{f}_t\\
	\mb{h}_t &= tanh(\mb{c}_t) \odot \mb{o}_t,
\end{align*}
where $\sigma(\cdot)$ is the sigmoid function.
The model's output at any time $t$ is given as a function of the hidden state $\mb{h}_t$. 

The accuracy of LSTM inference diminishes with increasing length of the time horizon. To alleviate that, we use DARNN that is an encoder-decoder version of LSTM with the attention mechanism, as seen in Figure~\ref{fig:arch}.  Herein, the inputs $\{\mb{x}_t\}$ can be transformed via an input attention mechanism that uses scaling to better ``highlight'' important parts of the input sequence. 
The process of generating multiplicative factors for the input attention mechanism (cf. Figure~\ref{fig:arch}) is delineated by the following equations.
\begin{subequations}
\label{eq:inpat}
\begin{align}
	\label{eq:inpatA}
	\mb{x}^i &= [x_{1,i}, \dots, x_{T,i}]^{\top}\\
	\label{eq:inpatB}
	\alpha_{t,i}' &= \texttt{InputAttnWts}\left(\mb{h}^e_{t-1},\mb{c}^e_{t-1},\mb{x}^i\right)\\
	\alpha_{t,i} &= \frac{\exp(\alpha_{t,i}')}{\sum_{j=1}^n \exp(\alpha_{t,j}')},\, i=1,2,\dots,n\\
	\label{eq:inpatD}
	\pmb{\alpha}_t &= [\alpha_{t,1},\dots,\alpha_{t,n}]^{\top},
\end{align}
\end{subequations}
where \texttt{InputAttnWts} consists of a fully connected multilayer perceptron (MLP). Then, the attention weight $\pmb{\alpha}_t$ is used to adjust the input features fed into the encoder LSTM, given as 
\begin{align}
	\label{eq:cell}
	(\mb{h}^e_t,\mb{c}^e_t) = \texttt{EncLSTMCell}(\pmb{\alpha}_t\odot \mb{x}_t,\mb{h}^e_{t-1},\mb{c}^e_{t-1}),
\end{align}
where $\mb{h}^e_t$ and $\mb{c}^e_t$ denote the hidden and cell states at time $t$, respectively. Additionally, $\mb{h}^e_t$ functions as the output of the encoder LSTM at time $t$. These outputs carry temporal information for the decoder. 

To perform a soft selection of the encoder outputs for the decoder LSTM input, a temporal attention mechanism is employed as follows.
\begin{subequations}
\label{eq:outat}
\begin{align}
	\label{eq:outatA}
	\beta_{t,t'}' &= \texttt{TempAttnWts}\left(\mb{h}^d_{t-1}\mb{c}^d_{t-1},\mb{h}^e_{t'}\right)\\
	\beta_{t,t'} &= \frac{\exp(\beta_{t,t'}')}{\sum_{s=1}^{T} \exp(\beta_{t,s}')}\\
	\label{eq:outatC}
	\pmb{\beta}_t &=[\beta_{t,1},\dots, \beta_{t,T}]^{\top},\,\, 
 \mathbf{H}^e = [\mb{h}^e_{1},\dots,\mb{h}^e_{T}]\\
	\label{eq:outatD}
	\tilde{y}_t &= \mb{w}_c^\top \begin{bmatrix}
\mathbf{H}^e \pmb{\beta}_t \\
y_t
\end{bmatrix} + b_c,
\end{align}
\end{subequations}
where \texttt{TempAttnWts} is a fully connected MLP. The context vector $\mathbf{H}^e \pmb{\beta}_t$ is a weighted sum of the hidden states. Equation~\eqref{eq:outat} comprises the temporal attention, concatenation, and projection blocks depicted in Figure~\ref{fig:arch}. The resulting scalar $\tilde{y}_t$ is used as input for the decoder LSTM at time $t$:
\begin{align}
	\label{eq:Dcell}
(\mb{h}^d_t,\mb{c}^d_t) = \texttt{DecLSTMCell}(\tilde{y}_t,\mb{h}^d_{t-1},\mb{c}^d_{t-1}).
\end{align}
To this end, we concatenate the decoder LSTM output $\mb{h}^d_T$ and the context vector at time $T$ and pass it through a fully connected MLP to generate the final forecast $\hat{y}_{T+L}$.
\begin{align}
	\label{eq:FC}
	\hat{y}_{T+L} =\texttt{FullyConnected}\left(\mb{h}^d_T,\mathbf{H}^e \pmb{\beta}_T\right).
\end{align}
Multiple LSTM layers can be stacked by utilizing the hidden state of lower layers as input for upper layers. In this context, a stacking size of 2 is employed for both the encoder and decoder LSTMs.

\subsubsection*{Data layout} 
The STLF model can be trained on the feature-target pairs $\{\mb{X}_d,Y_d\}_{d=1}^D$ with
\begin{align*}
	\mb{X}_d \ldef \{ \mb{x}_1,\dots,\mb{x}_{T},y_1,\dots,y_T \},\quad Y_d \ldef y_{T+L}.
\end{align*}
The training loss over a single sample can be simply the squared forecast error, which is defined as 
$$l(\hat{Y}_d,Y_d) = (\hat{Y}_d -Y_d)^2,$$ 
where the $L$-step ahead forecast value $\hat{Y}_d  = \texttt{DARNN}(\mb{X}_d)$ is the output of the DARNN model (cf. equations \eqref{eq:inpat}--\eqref{eq:FC}).

The contents of exogenous features $\mb{x}_t$ depend on the available data. For the present work, these features are described in Section~\ref{sec:simulations}. The model's learnable parameters include the weights $\mb{W}_{(\cdot)}$ and biases $\mb{b}_{(\cdot)}$ terms in the encoder and decoder LSTM cells, the \texttt{InputAttnWts} and \texttt{TempAttnWts}, the\texttt{FullyConnected} layers, as well as the projection parameters $\mb{w}_c$ and $b_c$. In the following sections, we use the symbol $\pmb{\theta}$ to denote the concatenated vector comprising all learnable parameters.

\section{Federated Learning Architecture with PLs}
\label{sec:plfl}

\begin{figure}[!tb]
	\centering
	\includegraphics[width=0.9\linewidth]{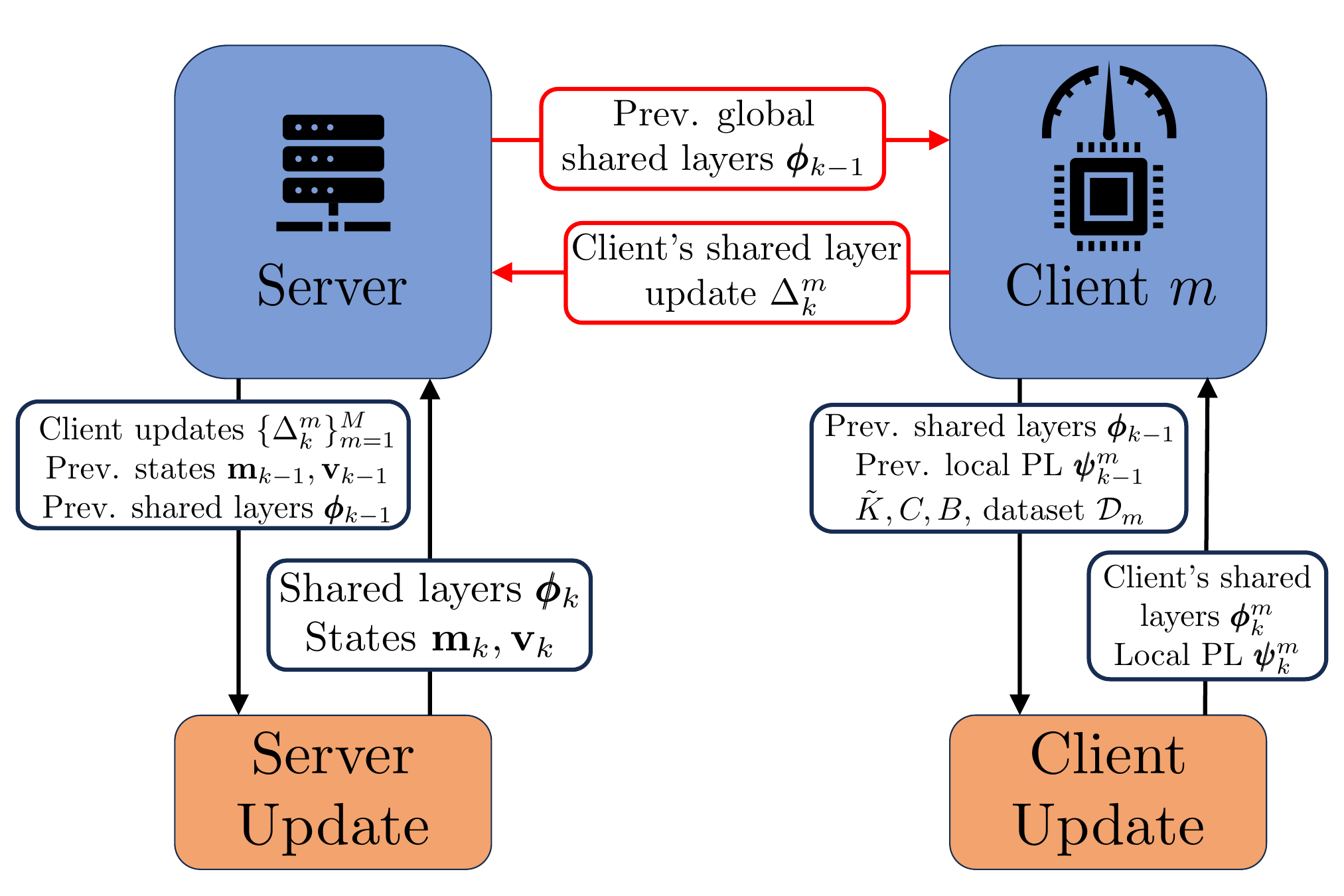}
	\caption{A schematic of PPFL. Communications and computation are represented by red and black arrows, respectively.}
	\label{fig:PLschm}
\end{figure}

\begin{algorithm}[!tb]
	\caption{Privacy-Preserving Federated Learning (PPFL)}
	\begin{algorithmic}[1]
		\renewcommand{\algorithmicrequire}{\textbf{Input:}}
		\renewcommand{\algorithmicensure}{\textbf{Output:}}
		\REQUIRE Datasets $\mcal{D}_m$ for clients $m\in[M]$, Server epochs $K$, Client epochs $\tilde{K}$, Update clip value $C>0$, minibatch size $B>0$, additional noise $\{\{\pmb{\xi}_{m,k}\}_{m=1}^M\}_{k=1}^K$
		\ENSURE Trained shared parameters $\pmb{\phi}_K$, client models $\{\bphi^m_K,\bpsi^m_K\}_{m=1}^M$\\
		\texttt{//Initialization of optimizer states}
		\STATE Initialize parameters $\pmb{\phi}_0$, states $\mb{m}_0,\mb{v}_0$ at server
		\STATE Initialize parameters $\{\bpsi^m_0\}$ at clients $m\in[M]$\\
		\texttt{//Training starts}
		\FOR {server epochs $k \in[K]$}
		\STATE Server sends $\pmb{\phi}_{k-1}$ to all clients
		\FOR{Client $m\in[M]$}
		\STATE $\pmb{\phi}^m_{k},\pmb{\psi}^m_{k} = \texttt{ClientUpdate}(\pmb{\phi}_{k-1},\pmb{\psi}_{k-1}^m,\mcal{D}_m,\tilde{K},C,B)$
		\texttt{//Client sends noisy shared layers  update}
		\STATE Client $m$ sends $\Delta^m_k \ldef \bphi^m_k - \bphi_{k-1} + \pmb{\xi}_{m,k}$ to server\label{lin:noisy}
		\ENDFOR
		\STATE $\bphi_k,\mb{m}_k,\mb{v}_k =$\\
		$ \texttt{ServerUpdate}(\bphi_{k-1},\{\Delta^m_k\}_{m\in[M]},\mb{m}_{k-1},\mb{v}_{k-1})$
		\ENDFOR
		\RETURN Server: $\bphi_K$, Clients: $\{\bphi^m_K,\bpsi^m_K\}_{m=1}^M$
	\end{algorithmic} 
	\label{alg:main}
\end{algorithm}

\begin{algorithm}[!tb]
	\caption{\texttt{ClientUpdate} with Adam}
	\begin{algorithmic}[1]
		\renewcommand{\algorithmicrequire}{\textbf{Input:}}
		\renewcommand{\algorithmicensure}{\textbf{Output:}}
		\REQUIRE Shared layer parameters $\bphi_{k-1}$, Client's PL parameters $\bpsi^m_{k-1}$, Client's dataset $\mcal{D}_m$, Client epochs $\tilde{K}$, Clip value $C$, Minibatch size $B$
		\ENSURE Updated shared parameters $\bphi^m_k$, PL parameters $\bpsi^m_k$
		\renewcommand{\algorithmicensure}{\textbf{Hyperparameters:}}
		\ENSURE $\beta_1,\beta_2\in(0,1)$, $\eta > 0$, $\delta > 0$
		\STATE Concatenate $\bt^m_{0,k} \ldef [\bphi_{k-1},\bpsi^m_{k-1}]$
		\STATE Initialize states $\hat{\mb{m}}_0^m=\mb{0}, \hat{\mb{v}}_0^m = \delta \mb{1}$ of same size as $\bt^m_{k-1}$
		\FOR{$\tilde{k} \in [\tilde{K}]$}
		\STATE Sample minibatch $\mcal{M}^m_{\tilde{k}}$ of size $B$ from $\mcal{D}_m$
		\STATE Get gradient $\mb{g}_{\tilde{k}} = \frac{1}{B}\nabla_{\bt} \sum_{(\mb{X},Y)\in\mcal{M}^m_{\tilde{k}}} l(f(\mb{X}|\bt^m_{\tilde{k}}),Y)$
		\STATE $\hat{\mb{m}}^m_{\tilde{k}} = \beta_1\hat{\mb{m}}^m_{\tilde{k}-1} + (1-\beta_1)\mb{g}_{\tilde{k}}$
		\STATE $\hat{\mb{v}}^m_{\tilde{k}} = \beta_2\hat{\mb{v}}^m_{\tilde{k}-1} + (1-\beta_2)\mb{g}_{\tilde{k}}^{(2)}$\\
		\texttt{//Local Adam updates}
		\STATE $\bt^m_{\tilde{k},k} = \bt^m_{\tilde{k}-1,k} - \eta \frac{  \hat{\mb{m}}^m_{\tilde{k}} (1-\beta_1^{\tilde{k}})^{-1}}{  \sqrt{  \hat{\mb{v}}^m_{\tilde{k}} (1-\beta_2^{\tilde{k}})^{-1}  } + \delta\mb{1}   }$
		\ENDFOR
		\texttt{\\//Clip L1 norm of update to C}
		\IF{$\left\lVert \bt^m_{\tilde{K},k} - \bt^m_{0,k} \right\rVert_1 > C$} \label{Clientline}
		\STATE $\bt^m_{\tilde{K},k} = \bt^m_{0,k} + C \frac{ \bt^m_{\tilde{K},k} - \bt^m_{0,k} }{ \left\lVert \bt^m_{\tilde{K},k} - \bt^m_{0,k} \right\rVert_1 }$\label{line:clip}
		\ENDIF
		\STATE Extract $[\bphi^m_k,\bpsi^m_k]$ from $\bt^m_{\tilde{K},k}$
		\RETURN $\bphi^m_k,\bpsi^m_k$
	\end{algorithmic} 
	\label{alg:client}
\end{algorithm}

\begin{algorithm}[!tb]
	\caption{\texttt{ServerUpdate} with FedAdam}
	\begin{algorithmic}[1]
		\renewcommand{\algorithmicrequire}{\textbf{Input:}}
		\renewcommand{\algorithmicensure}{\textbf{Output:}}
		\REQUIRE Previous shared layer parameters $\pmb{\phi}_{k-1}$, Shared layer updates $\{\Delta^m_k\}_{m=1}^M$, Server optimizer states $\mb{m}_{k-1}, \mb{v}_{k-1}$
		\ENSURE Updated shared parameters $\bphi_k$, states $\mb{m}_k,\mb{v}_k$
		\renewcommand{\algorithmicensure}{\textbf{Hyperparameters:}}
		\ENSURE $\bar{\beta}_1,\bar{\beta}_2\in(0,1)$, $\bar{\eta} > 0$, $\bar{\delta} > 0$\\
		\texttt{//Average updates from all clients}
		\STATE $\Delta_k = \frac{1}{M} \sum_{m=1}^M \Delta^m_k$\\
		\STATE $\mb{m}_k = \bar{\beta_1} \mb{m}_{k-1} + (1-\bar{\beta}_1)\Delta_k$
		\STATE $\mb{v}_k = \bar{\beta}_2 \mb{v}_{k-1} + (1-\bar{\beta}_2)\Delta_k^{(2)}$\\
		\texttt{//Server FedAdam update}
		\STATE $\bphi_k = \bphi_{k-1} + \bar{\eta} \frac{ \mb{m}_t }{\sqrt{\mb{v}_t} + \bar{\delta}}$
		\RETURN $\bphi_k,\mb{m}_k,\mb{v}_k$
		\label{line:PLServerEnd} 
	\end{algorithmic} 
	\label{alg:server}
\end{algorithm}

Training the STLF model involves learning a mapping of the form $Y = f(\mb{X}|\pmb{\theta})$ wherein $f(\cdot)$ is the DARNN model. We assume that $\bt$ can be split as $\bt = [\bphi,\bpsi]$ where $\bphi$ and $\bpsi$ represent the parameters of the shared layers and personalized layers, respectively.  In the setting of federated learning, we consider $M$ clients, each with a local dataset $\mcal{D}_m \ldef \{\mb{X}_d,Y_d\}_{d\in[D_m]}$ consisting of $D_m$ data points. Then, FL with PLs aims to minimize the following empirical loss:
\begin{align}
	\label{eq:opt}
	\min\limits_{\bphi,\{\bpsi\}_{m=1}^M} \;\; \frac{1}{M} \sum_{m=1}^M \frac{1}{|\mcal{D}_m|}  \sum_{(\mb{X},Y)\in\mcal{D}_m}l\left(f(\mb{X}|\bphi,\bpsi_m\right),Y).
\end{align}

PL provides each client with its personalized STLF model, which can be evaluated locally to provide forecasts as needed. We solve 
\eqref{eq:opt} by using the proposed PPFL, as presented in Algorithm~\ref{alg:main}.
At each server epoch $k$, PPFL maintains a global copy of the shared parameter $\bphi_k$, which is broadcast to all clients. Then, each client combines it with their locally stored PL parameter $\bpsi_k^m$ to form the complete weight $\bt^m_k$. 
Subsequently, each client executes $\tilde{K}$ rounds of Adam~\cite{ADAM} to update $\bt^m_k$ using their local data (cf.~Algorithm~\ref{alg:client}). 
Finally, all clients feed the updates of their shared layers $\Delta_k^m$ back to the server, which aggregates $\{\Delta_k^m\}_{m=1}^M$ via FedAdam~\cite{FEDADAM} to update the shared layers (cf.~Algorithm~\ref{alg:server}). FedAdam replicates the dynamics of Adam, including the maintenance of global states $\mb{m}$ and $\mb{v}$, for the aggregation of all model updates.

It is important to acknowledge that, although the pairing of Adam and FedAdam yields favorable outcomes for the present work, alternative client-side algorithms like stochastic gradient descent (SGD), Adagrad, RMSprop, and server-side algorithms such as FedSGD, FedAdagrad, etc, are also viable options (see e.g.,~\cite{FEDADAM}).

Two crucial aspects of Algorithm~\ref{alg:main} empower the implementation of model obfuscation. The first is the inherent obfuscation offered by personalized layers (PLs), as they are not accessible to the server and other clients, preventing a complete model evaluation. The second involves the clients' ability to transmit noisy variants of shared parameter updates to the server (Algorithm~\ref{alg:main}, line~\ref{lin:noisy}). By selecting carefully calibrated noise following the principles of differential privacy (DP), we can safeguard the privacy of information contained in the updates, thereby enhancing privacy protection.

The core principle of DP in a generic setting involves adding a pre-calibrated noise to the output of a function or mechanism, which makes it impossible to distinguish between similar inputs up to a certain probability. More concretely, consider a (possibly randomized) function $\mcal{F}(\mb{x})$, and let $\mb{x} \sim_1 \bar{\mb{x}}$ imply that $\mb{x}$ and $\bar{\mb{x}}$ differ in at most one entry. If it holds for any $S \subseteq \Ima(\mcal{F})$ and $\mb{x}\sim_1\bar{\mb{x}}$ such that
\begin{align*}
	\frac{\text{Pr}\left[ \mcal{F}(\mb{x}) \in S \right]}{\text{Pr} \left[ \mcal{F}(\bar{\mb{x}}) \in S\right]} \leq \exp(\epsilon),
\end{align*}
then $\mcal{F}$ is called $\epsilon$-DP. The amount of privacy afforded by an $\epsilon$-DP mechanism is inversely proportional to the \emph{privacy budget} $\epsilon >0$. 
For a vector-valued function $\mathbf{f}(\cdot)$ that satisfies
$\max\limits_{\mb{x}\in \text{dom}(\mathbf{f})} \norm{\mathbf{f}(\mb{x})} \leq C$, the mechanism $\mathbf{f}(\mb{x}) + \boldsymbol{\xi}$ is $\epsilon$-DP, where $\boldsymbol{\xi}$ is a random vector with independent and identically distributed (i.i.d.) elements $\boldsymbol{\xi}_i\sim \text{Laplace}\left(0,\frac{2C}{\epsilon} \right)$ for all $i$.

\begin{lemma}
	If $\boldsymbol{\xi}_{m,k}$ is elementwise i.i.d. as $\text{Laplace}\left(0,\frac{2C}{\epsilon}\right)$ for all $m\in[M]$ and $k\in[K]$, then client's update to server, i.e. $\Delta^m_k$ is $\epsilon$-DP with respect to the true update $\bphi^m_k-\bphi_{k-1}$.
\end{lemma}

The proof follows immediately from the fact that the true updates $\bphi^m_k-\bphi_{k-1}$ are clipped to at most $C$ (cf.~Algorithm~\ref{alg:client}, line~\ref{line:clip}) in the sense of $\ell_1$-norm. In the following section, we will explore model accuracy for different values of $\epsilon$.

\section{Simulation Results}
\label{sec:simulations}


\begin{figure}[!tb]
	\centering
	\includegraphics[width=\linewidth]{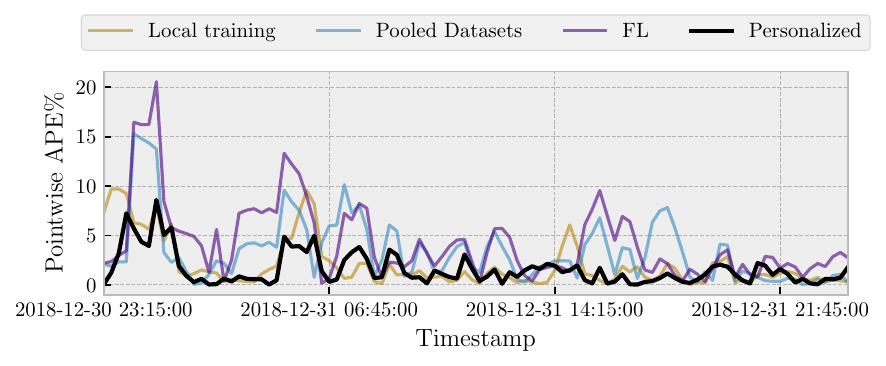}
	\caption{Forecasting with models trained with different methods. Plotted are the last 100 data points from Client 1's test set.}
	\label{fig:test}
\end{figure}

\begin{figure}[!tb]
	\centering
	\includegraphics[width=\linewidth]{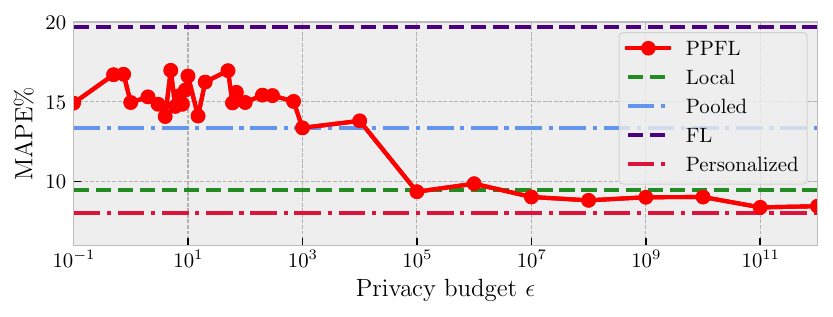}
	\caption{MAPE errors for clients trained with PPFL for different values of privacy budget $\epsilon$. The errors are averaged across the test sets of all 8 clients.}
	\label{fig:mape}
\end{figure}

We use the NREL ComStock dataset that contains load data of the commercial US building stock over 2018~\cite{COMSTOCK}. The dataset contains multiple exogenous features such as building characteristics, weather information, etc. The feature vector $\mb{x}_t$ is given as
$$
	\mb{x}_t = \begin{bmatrix*}[l]
		\text{15-min interval index within a day at $t$ ($\{0,\dots,95\}$)} \\
		\text{day of week index at $t$ ($\{0,\dots,6\}$) }\\
		\text{global horizontal radiation at $t$}\\
		\text{temperature at $t$}\\
		\text{wind speed at $t$}\\
		\text{areas of building floor, window, and roof}\\
		\text{cooling equipment capacity}
	\end{bmatrix*}.
$$

We select eight distinct buildings (clients) in New York, USA, characterized by substantial heterogeneity in both the scale and variance of loads across time (cf.~Table~\ref{tab:0}). 
We train the DARNN model using those clients' data. Subsequently, we assess the performance of PPFL for various values of the privacy budget $\epsilon$ by configuring the encoder layers as shared and the decoder layers as personalized (PL), as illustrated in Figure~\ref{fig:arch}.
Finally, we compare the proposed PPFL with the following non-DP schemes.
\begin{itemize}
	\item \textbf{Local}: Each client trains a local STLF model exclusively on its own data with Adam.
	\item \textbf{Pooled}: All clients' data are pooled into a single dataset, on which a single model is trained with Adam.
	\item \textbf{FL}: Classical federated learning, which is PPFL where all layers are shared and no noise.
	\item \textbf{Personalized}: Each client trains its own model with PPFL by marking the decoder as PL and no noise.
\end{itemize}

\begin{table}[!tb]
	\centering
	\caption{Mean and variance of loads across time.}
	\begin{tabular}{c|c|c}
		\hline
		\textbf{Client \#} & \textbf{Mean (kWh)} & \textbf{Variance}\\
		\hline
		 1 & 488.04 & 10298.24\\
		 2 & 204.59 & 21119.52\\
		 3 & 176.54 & 3505.70\\
		 4 & 156.63 & 2780.87\\
		 5 & 107.12 & 5314.93\\
		 6 & 59.18 & 1906.06\\
		 7 & 42.32 & 888.42\\
		 8 & 22.08 & 137.23\\
		\hline
	\end{tabular}
	\label{tab:0}
\end{table}

\begin{table}[!tb]
	\centering
	\caption{MASE and MAPE errors for model trained by various methods. The errors are averaged across all 8 clients.}
	\begin{tabular}{l|c|c}
		\hline
		\textbf{Method} & \textbf{MASE} & \textbf{MAPE}\\
		\hline
		\textbf{Local} & 0.528 & 9.46\%\\
		\textbf{Pooled} & 0.827 & 13.36\%\\
		\textbf{FL} & 1.125 & 19.67\%\\
		\textbf{Personalized} & \textbf{0.477} & \textbf{8.01\%} \\
            \textbf{PPFL} ($\epsilon=10000$) & 0.584 & 9.36\%\\
		\textbf{PPFL} ($\epsilon=1000$) & 0.761 & 13.37\%\\
            \textbf{PPFL} ($\epsilon=100$) & 0.896 & 14.96\%\\
		\textbf{PPFL} ($\epsilon=10$) & 0.960 & 16.62\%\\
		\textbf{PPFL} ($\epsilon=1$) & 0.851 & 14.96\%\\
		\textbf{PPFL} ($\epsilon=0.1$) & 0.822 & 14.92\%\\
		\hline
	\end{tabular}
	\label{tab:1}
\end{table}

The dataset has a temporal granularity of 15 minutes. We set $T=12$ and $L=4$, i.e. 
predicting the next-hour load demand using the data points from the last three hours. 
We employ a batch size of 64 for all experiments, except for the pooled method, where the batch size is increased eightfold (i.e. $512$), to ensure consistency in the effective number of updates per client across different methods. We set $\beta_1=0.9, \beta_1=0.999$ for the clients and $\bar{\beta_1}=0.99,\bar{\beta}_2=0.999$ for the server. The adaptivity parameters $\delta$ and $\bar{\delta}$ are set to \texttt{1e-8} while $\eta$ and $\bar{\eta}$ are set to 0.001 and 0.01, respectively. For DARNN, we choose the architecture from Figure~\ref{fig:arch} with two LSTM layers and states $\mb{h},\mb{c}\in\real{30}$ for both the encoder and decoder. All MLPs are two-layered with tanh activation for attention weights and ReLU activations otherwise. All methods are trained with $K=4000$ and $\tilde{K}=5$ except for the pooled training, which is trained with $\tilde{K}K=20000$ epochs. For all methods, the clip value is set to $C=200$. The data for each client are divided into training, validation, and test sets in an 80:10:10 ratio along the time axis. The codes are written using PyTorch and Advanced Privacy-Preserving Federated Learning (APPFL) package~\cite{ryu2022appfl}.

Given a time series $\{y_t\}_{t=1}^{T_{f}}$ and its forecast $\{\hat{y}_t\}_{t=1}^{T_{f}}$, we utilize two error metrics: mean absolute scaled error (MASE) and mean absolute percentage error (MAPE). These metrics are defined as:
\begin{align*}
	\text{MASE} &= \frac{\sum_{t\in \mathcal{T}} |y_{t} - \hat{y}_t|}{ \sum_{t\in \mathcal{T}} |y_t - y_{t-L}|},\\
	\text{MAPE} &= \frac{1}{|\mathcal{T}|}\sum_{t\in \mathcal{T}} \frac{|y_{t} - \hat{y}_t|}{|y_{t}|} \times 100\%,
\end{align*}
where $\mathcal{T}\triangleq \{t : t=T+L+1,\dots, T_{f}\}$ is the testing horizon.
Figure~\ref{fig:test} displays the pointwise absolute percentage error (APE) values for the non-DP cases on the test set of client 1. The average MASE and MAPE errors of all 8 clients are provided in Table~\ref{tab:1}. It should be noted that if a forecasting method yields a MASE value exceeding one, it is considered inferior to persistence forecasting, which naively repeats the last known data point.

The personalized model without differential privacy achieves the best performance on both metrics. This outcome underscores the advantages of personalized layers (PLs) compared to classical federated learning, local training, and single models trained on a pooled dataset.
On the other hand, PPFL demonstrates commendable results with budgets $\epsilon = 10^{p}$ for $p=-1,0,\dots,4$. Intuitively, as $\epsilon\rightarrow \infty$, PPFL should approach the performance of non-DP personalization. This is confirmed in Figure~\ref{fig:mape} by comparing average MAPE errors for various values of $\epsilon$. The results illustrate that PPFL is capable of training models that perform well even under substantial differential privacy noise (e.g., $\epsilon<1$). Ultimately, the tuning of $\epsilon$ involves considering tradeoffs between privacy level and model accuracy.

\section{Conclusion}

In this paper, we demonstrate the utility of personalization in addressing the diminished performance of federated learning in the presence of heterogeneous clients. Additionally, we introduce the PPFL algorithm designed to safeguard the privacy of shared layer parameters through differential privacy. The simulation results provide validation for the effectiveness of the proposed method. Future research directions include conducting a convergence analysis of the proposed method and exploring the impact of utilizing forecasts generated by PPFL with obfuscation on downstream applications, such as voltage control.

\bibliographystyle{IEEEtran}
\bibliography{refs.bib}

\end{document}